\documentclass[aps,prl,superscriptaddress,floats,showpacs,floatfix, reprint]{revtex4-1}

\usepackage{amsmath}
\usepackage{amssymb}
\usepackage{ulem}

\usepackage{xcolor}
\usepackage{graphicx}
\usepackage{subfigure}
\usepackage{bm}
\usepackage{hyperref}

\begin{document}

\title{Origin of the Coulomb Pseudopotential}

\author{Tao Wang}
\affiliation{Department of Physics, University of Massachusetts, Amherst, Massachusetts 01003, USA}
\author{Xiansheng Cai}
\affiliation{Department of Physics, University of Massachusetts, Amherst, Massachusetts 01003, USA}
\author{Kun Chen}
\email{kunchen@flatironinstitute.org}
\affiliation{Center for Computational Quantum Physics, Flatiron Institute, 162 5th Avenue, New York, New York 10010}
\author{Boris V. Svistunov}
\email{svistunov@physics.umass.edu}
\affiliation{Department of Physics, University of Massachusetts, Amherst, Massachusetts 01003, USA}
\affiliation{Wilczek Quantum Center, School of Physics and Astronomy, Shanghai Jiao Tong University, Shanghai 200240, China}
\author{Nikolay V. Prokof'ev}
\email{prokofev@physics.umass.edu}
\affiliation{Department of Physics, University of Massachusetts, Amherst, Massachusetts 01003, USA}

	
\date{\today}
\begin{abstract}
   We address the outstanding problem of electron pairing in the presence of strong Coulomb repulsion at small to moderate values of the Coulomb parameter, $r_s \lesssim 2$, and demonstrate that the pseudopotential framework is fundamentally biased and uncontrolled. Instead, one has to break the net result into two distinctively different effects: 
   the Fermi liquid renormalization factor and the change in the effective low-energy coupling. The latter quantity is shown to behave non-monotonically with an extremum at
	$ r_s\approx 0.75$. Within the random-phase approximation,  Coulomb interaction starts to {\it enhance} the effective pairing coupling at $r_s >2$, and the suppression of the critical temperature is entirely due to the renormalized Fermi liquid properties.   
	Leading vertex corrections change this picture only quantitatively. 
	Our results call for radical reconsideration of the widely accepted repulsive pseudopotential approach and show the need for precise microscopic treatment
	of Coulomb interactions in the problem of superconducting instability.
\end{abstract}

\maketitle
\textit{Introduction.}\textemdash
The paring of electrons in the presence of strong repulsive Coulomb forces remained 
unsolved for nearly half a century until it was recognized that in the vast majority of low-temperature superconductors, the scenario of Cooper instability
is of the emergent BCS type, implying a quantitatively accurate low-energy effective description in terms of the two (partially related) parameters: the energy-frequency cutoff $\omega_0 \ll E_F$ ($E_F$ is the Fermi energy; $\hbar=1$) and the dimensionless effective coupling constant $g\ll 1$. Within this effective BCS theory, the expression for the critical temperature reads
\begin{equation}
T_c \, = \, \omega_0  e^{-1/g} \, .
\label{T_c}
\end{equation}
For the phonon-mediated Cooper instability, one has $\omega_0 \sim \omega_{\rm ph}$, where $ \omega_{\rm ph}$ is a typical phonon frequency. (The exact choice of $\omega_0$ is a matter of convention, because changes in $\omega_0$ can be absorbed into $g$.) 

The emergent BCS regime implies that $g$ can be decomposed into a product of two distinctive factors---the pseudopotential $U$ and the Fermi liquid factor $f_{\rm FL}$:
\begin{equation}
g \, \propto\,U f_{\rm FL}\, .
\label{g_U_f}
\end{equation}
The pseudopotential is understood as an amplitude of the dimensionless attractive coupling between {\it bare} electrons near the Fermi surface (FS), and $f_{\rm FL}$ is given by
\begin{equation}
f_{\rm FL}\, = \, z^2 \, (m_*/m_0) \, ,
\label{f_FL}
\end{equation}
where $z$ is the quasiparticle residue and $m_*/m_0$ is the FS 
effective mass renormalization.
It accounts for the fact that we are dealing with the correlated 
liquid rather than an ideal gas.  
Exponential sensitivity of the critical temperature to the small parameter 
$g$ implies that the positive-definite factor $f_{\rm FL}$---if noticeably smaller than unity---can dramatically suppress the value of $T_c$.

The strength of Coulomb interaction is characterized by the dimensionless parameter (the Wigner-Seitz radius) 
$ r_s= [ (4\pi /3) a_0^3 n ]^{-1/3} $,
where $n$ is the number density and $a_0$ is the Bohr radius. Typical experimental values of $r_s \gtrsim 2$ correspond to a moderately strong interaction. 
{\it A priori} one expects that Coulomb repulsion 
simply eliminates the possibility of phonon-mediated pairing in materials, but  experiment tells us otherwise.
The Coulomb pseudopotential framework, developed in the late 1950s \cite{Tolmachev1961,morelCalculationSuperconductingState1962a}, offers an empirical method to account for Coulomb interactions in superconductors. It has been successfully applied to estimate $T_c$ in a large number of experiments by means of a semi-phenomenological fitting procedure based on McMillan's formula \cite{mcmillanTransitionTemperatureStrongCoupled1968,calandraTheoreticalExplanationSuperconductivity2005,hiroseElectronicStatesMetallic2022,lianTwistedBilayerGraphene2019,naskarExperimentalFirstprinciplesStudies2022,tanChargeDensityWaves2021}. The framework, however, only provided a limited understanding because it neglected (i) the dynamic nature of screening in metals, (ii) renormalization of single-particle properties, and (iii) changes in the frequency and momentum dependence 
of the gap function when different mechanisms are combined. These conceptual mistakes prevent the development of better methods for material science calculations, and therefore this framework needs to be replaced with controlled first-principles treatments.

By accounting only for logarithmic suppression of the 
frequency-independent repulsion near the FS, Refs.~\cite{Tolmachev1961,morelCalculationSuperconductingState1962a}
argued that the net effect can be reduced to the 
so-called repulsive Coulomb pseudopotential
\[
\mu^* \, = \, \frac{\mu}{1 + \mu \ln (E_F/\omega_0) } \, ,
\]
with $\mu > 0$ introduced in a rather uncontrolled fashion as a
coupling constant characterizing screened interaction 
(if $\mu$ is computed from $ \rho_0 4\pi e^2 / \kappa^2 $,
where $\rho_0$ is the ideal gas FS density of states 
per spin component and $\kappa$ is the Thomas-Fermi momentum, 
then $\mu = 0.5$).
The main effect of $\mu^*$ is to reduce the magnitude of the 
phonon-mediated $U$ as $U \to U - \mu^*$, with most experiments suggesting that $\mu^* \in (0.1 \div 0.15)$. The
small values of $\mu^*$ are explained by the large $E_F-to-\omega_0$ ratio, 
but neither its value nor its sign is derived from first principles, not to mention that Coulomb repulsion cannot be fully screened at finite frequency. 

\begin{figure}[htbp]
	\begin{center}
	\includegraphics[width=1.0\linewidth]{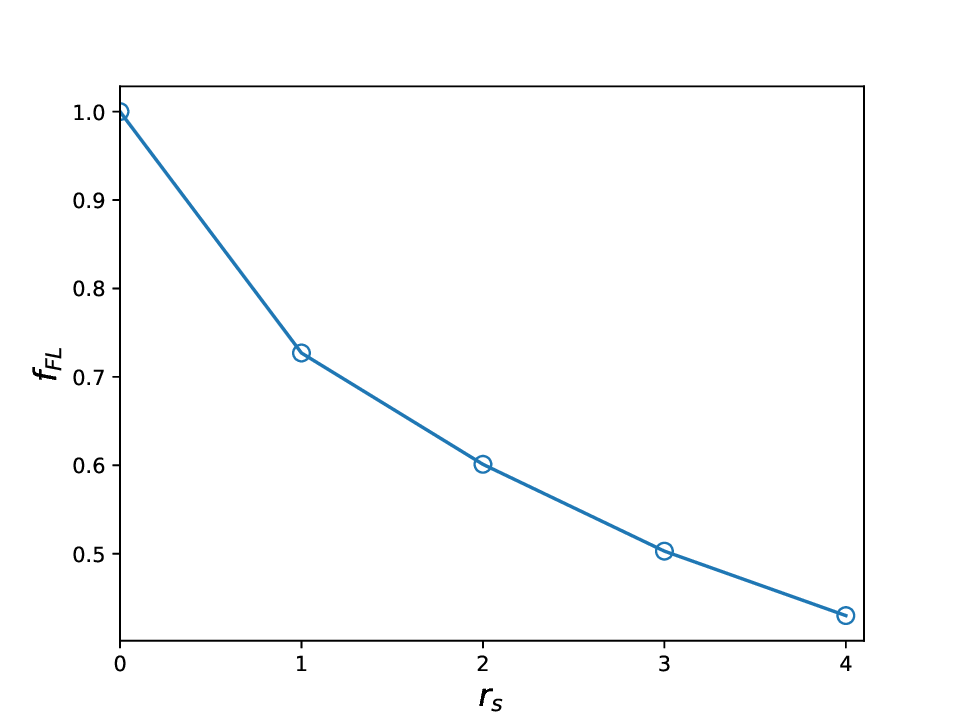}
	\end{center}
	\caption{
	\label{fig:f_FL}
	Fermi liquid factors $f_{\rm FL}$ of the uniform electron gas computed using
	data reported in Ref.~\cite{hauleSingleparticleExcitationsUniform2022}.
	}
\end{figure}

A recent breakthrough in the precise computation of the Fermi liquid properties 
of a uniform electron gas \cite{hauleSingleparticleExcitationsUniform2022} 
establishes that $f_{\rm FL}$ is significantly smaller than unity at $r_s > 2$ (see Fig. \ref{fig:f_FL}), in direct contradiction with the pseudopotential description,
see also Ref.~\cite{akashiRevisitingHomogeneousElectron2022} 
for the random-phase approximation (RPA) results in the same context.
To reconcile this finding with the experimental fact that corrections
to $g$ are small, one is forced to reconsider the effect 
of Coulomb potential on $U$---it has to be far smaller 
than predicted by $\mu^*$ and possibly even opposite in sign, 
i.e. Coulomb interactions in the $s$-wave channel 
might actually {\it increase} the amplitude of attractive $U$!

\begin{figure}[t]
	\begin{center}
	\includegraphics[width=1.0\linewidth]{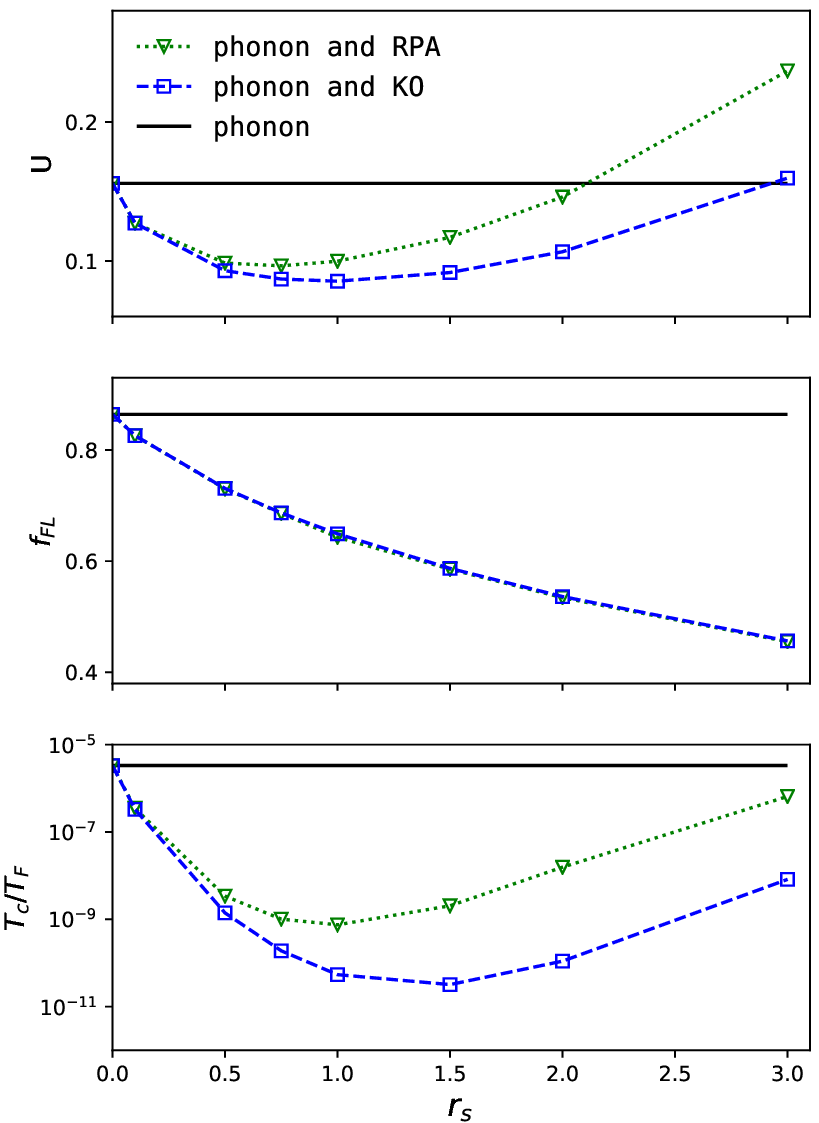}
	\end{center}
	\caption{\label{fig:main} Effective couplings, Fermi liquid factors, 
	 and critical temperatures for phonon-mediated superconductivity with 
	 and without the Coulomb vertex function approximated by either the RPA or Kukkonen-Overhauser (KO) interaction. Both approximation lead to qualitatively similar results for $U$: 
	 As $r_s$ increases, the effective coupling goes through a minimum and starts to
	 increase. The critical temperature follows a similar trend.}
\end{figure}

In this Research Letter, we employ an implicit renormalization protocol and a generalized discrete Lehmann representation for extracting the
effective coupling constant and critical temperature from the gap 
function equation \cite{chubukovImplicitRenormalizationApproach2019a}
to study the effect of
Coulomb repulsion on $U$ and $T_c$ (see Fig. \ref{fig:main}). We account for both the single-particle properties and the dynamic nature of screening with (i) dynamically screened Coulomb vertex functions, the use of which guarantees quantitative accuracy at $r_s \lesssim 2$; (ii) a fine, non-uniform momentum grid that resolves sharp behaviors near the Fermi surface and a frequency grid that covers a frequency range much larger than $E_F$ and (iii) a consistently renormalized Green's function based on the self-energies emerging from the same vertex function used in the gap equation.
We reveal that the suppression of $U$ is maximal at $r_s \approx 0.75$, 
and the effect diminishes for larger values of $r_s$. Within the RPA, Coulomb interactions start to \textit{enhance} attractive coupling at $r_s > 2$, but this result is sensitive to inclusion of vertex corrections. 
We discuss our findings in the context of earlier work suggesting or pointing 
to a possibility of pairing instability
in the absence of electron-phonon coupling, i.e. exclusively on the basis of dynamically screened Coulomb repulsion \cite{Tolmachev1961,takadaPlasmonMechanismSuperconductivity1978,rietschelRoleElectronCoulomb1983,richardsonEffectiveElectronelectronInteractions1997b}. Our results demonstrate an unambiguous separation of different effects of Coulomb interaction, disproving the idea of absorbing all of them into a single effective parameter---the pseudopotential.

\textit{Model.}\textemdash
The Hamiltonian of the uniform electron gas (UEG) on a neutralizing
background is defined as
\begin{equation}
  \label{eq:hamiltonian}
  H \! = \! \sum\limits_{\vec{k} \sigma} \epsilon_{\vec{k}} a_{\vec{k} \sigma}^\dagger a_{\vec{k} \sigma}
  + \frac{1}{2}\sum\limits_{\vec{q}\neq 0}\sum\limits_{\vec{k} \sigma}\sum\limits_{\vec{k^\prime} \sigma^\prime}
  V_{q} a_{\vec{k}+\vec{q} \sigma}^\dagger a_{\vec{k^\prime}\! -\vec{q} \sigma^\prime}^\dagger a_{\vec{k^\prime} \sigma^\prime}a_{\vec{k} \sigma},
\end{equation}
Here $a_{\vec{k} \sigma}^\dagger$ is the creation operator of an electron with momentum $\vec{k}$ and spin $\sigma=\uparrow,\downarrow$,
$\epsilon_{k}=\frac{k^2}{2m_0}-\mu$, and 
$V_{q}= \frac{4\pi e^2}{q^2}$ is the bare Coulomb interaction.

The gap function equation in the singlet channel reads
\begin{equation}
\label{eq:gap}
\lambda \, \Delta_{\omega_n,\mathbf{k}}
=
-T\sum\limits_{m}\int\frac{d\mathbf{p}}{{(2\pi)^d}}
\Gamma_{\omega_m,\mathbf{p}}^{\omega_n,\mathbf{k}} G_{\omega_m,\mathbf{p}}G_{-\omega_m,-\mathbf{p}}\Delta_{\omega_m,\mathbf{p}}.
\end{equation}
Here $\Gamma$ is the particle-particle irreducible four-point vertex, $G$ is the
single-particle Green's function, $\Delta$ is the gap function, and $\lambda\equiv \lambda(T) $ is its eigenvalue. The critical temperature $T_c$ corresponds to the point where 
$\lambda_{\rm max}(T)=1$.  

We consider two approximations for $\Gamma$ based on the the screened Coulomb interaction, both depending only on the momentum and energy transfer,
$\Gamma_{\omega_m,\mathbf{p}}^{\omega_n,\mathbf{k}} = \Gamma (\omega_m-\omega_n, \mathbf{p}-\mathbf{k}) $.
The RPA form is standard:
$\Gamma_{\rm RPA} = [V_q^{-1} + \Pi_0(\omega,q)]^{-1}$,
where $\Pi_0$ is the polarization function computed from the convolution of the bare Green's function. For simplicity we take the functional form of $\Pi_0$ to be that at $T=0$, which is justified by the smallness of the critical temperature.
To account for vertex corrections and estimate their role 
as a function of $r_s$, we employ the Kukkonen-Overhauser ansatz \cite{kukkonenElectronelectronInteractionSimple1979a}  when
\begin{equation}
\Gamma_{\rm KO}= V_q + V_+(q)^2 Q_+(\omega,q)
 -3V_-(q)^2 Q_-(\omega,q),
\end{equation}
with
$Q_\pm(\omega,q)=-[\Pi_0^{-1}(\omega,q) +V_\pm(\omega,q)]^{-1}$ and 
$V_+=(1-G_+)V_q$, $V_-=-G_-V_q$. 
Here $\Gamma_{\rm KO}$ is already projected on spin-singlet state as required by the fermionic parity. The higher-order vertex corrections neglected in the RPA are encoded in the local field factors $G_\pm(q)$ for which we adopt the ansatz proposed by Takada  \cite{takadaSuperconductivityOriginatingRepulsive1989}.

Finally, we introduce phonon-mediated interactions taken to have the same functional form as considered by Richardson and Ashcroft to study the very same problem of superconductivity in the UEG with electron-phonon coupling \cite{richardsonEffectiveElectronelectronInteractions1997b}.  
\begin{equation}
\label{eq:elph}
\Gamma_{\rm ph}(\omega , q) = -\frac{a \rho_0}{1+{(q/2k_F)}^2}\frac{\omega_q^2}{\omega^2+\omega_q^2} \, ,	
\end{equation}
with the phonon dispersion $\omega_q^2 = \frac{{\omega_{\rm ph}}^2(q/k_F)^2}{1+(q/k_F)^2}$ and dimensionless coupling strength $a$. For every choice of the vertex function
considered in this Research Letter the single-particle self-energy was computed 
self-consistently from the convolution of $G$ and $\Gamma $. 

\textit{Implicit renormalization approach.}\textemdash For the simplest 
case when $\Gamma = \Gamma_{\rm ph}$, the eigenvalue $\lambda(T)$ is a linear function of $\ln T$ at low temperature $T\ll\omega_{\rm ph}$ that can be 
written as
\begin{equation}
\label{eq:BCSflow}
\lambda(T) = - g \ln ( T / \omega_{0} ) \,. 	
\end{equation}
As expected, the condition $\lambda_{\rm max} (T_c)=1$ leads to Eq.\eqref{T_c}, 
and $T_c$ can be determined accurately by fitting the data even if 
calculations need to be stopped at $T\gg T_c$.
When Coulomb interactions are included, screening and renormalization effects
taking place in a broad frequency range above the phonon frequency ensure that
$\lambda_{\rm max} (T)$ is an unknown non-linear function of $\ln T$ that can be used
neither for reliable extrapolation towards lower temperature nor for evaluation of the
effective low-energy coupling $U$. The implicit renormalization (IR) 
approach of Ref.~\cite{chubukovImplicitRenormalizationApproach2019a} provides a solution to both problems by formulating an alternative eigenvalue problem.
The gap function is decomposed into two complementary (low-frequency and high-frequency)  parts, $\Delta=\Delta^{(1)}+\Delta^{(2)}$, with $\Delta_n^{(1)}=0$ for $|\omega_n|>\Omega_c$, and $\Delta_n^{(2)}=0$ for $|\omega_n|<\Omega_c$, and
the eigenvalue problem is solved for $\Delta_n^{(1)}$ only. The condition 
$\bar{\lambda}(T_c)=1$ for the largest eigenvalue of the new problem 
remains exact.

\begin{figure}[htbp]
	\centering
	\includegraphics[width=1.0\linewidth]{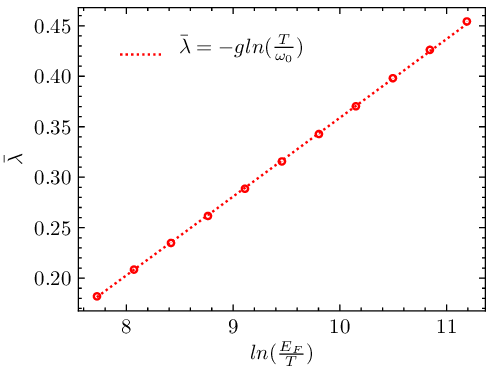}
	\caption{\label{fig:flow} Temperature dependence of the largest eigenvalue $\bar{\lambda}$ for $\Gamma = \Gamma_{\rm ph}+\Gamma_{\rm RPA}$ at $r_s = 2$. The emergent BCS linear flow with effective coupling constant $g$ and energy scale $\omega_{0}$ is represented by the dotted line.}
\end{figure}

As shown in Fig.~\ref{fig:flow}, the IR formulation brings back a nearly perfect linear dependence of $\bar{\lambda}$ on $\ln T$ for a properly chosen frequency scale separation $\Omega_c$. The slope of the curve is the emergent low-frequency
coupling strength $g$, while the vertical axis intercept determines the characteristic 
low-frequency scale $\omega_{0}$. Linear dependence is also crucial for accurate determination of $T_c$ from simulations performed at $T \gg T_c$ when $T_c$ is extremely low and the number of Matsubara frequency points required to solve the gap equation is large 
(this is done efficiently by the generalized discrete Lehmann representation  \cite{kayeDiscreteLehmannRepresentation2022, SDLR}, see Supplemental Material\cite{supplement}).

\begin{figure*}[htbp]
	\centering
	\includegraphics[width=0.47\linewidth]{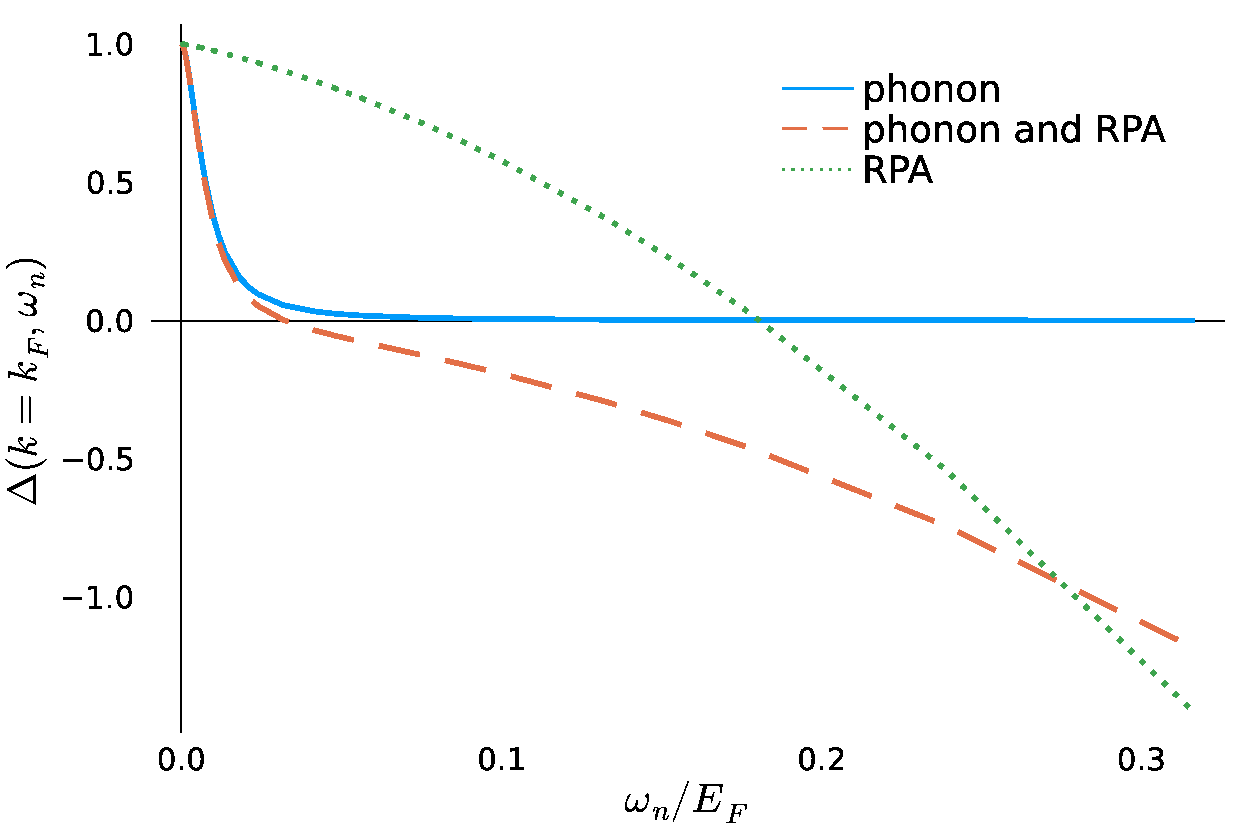}
	\includegraphics[width=0.47\linewidth]{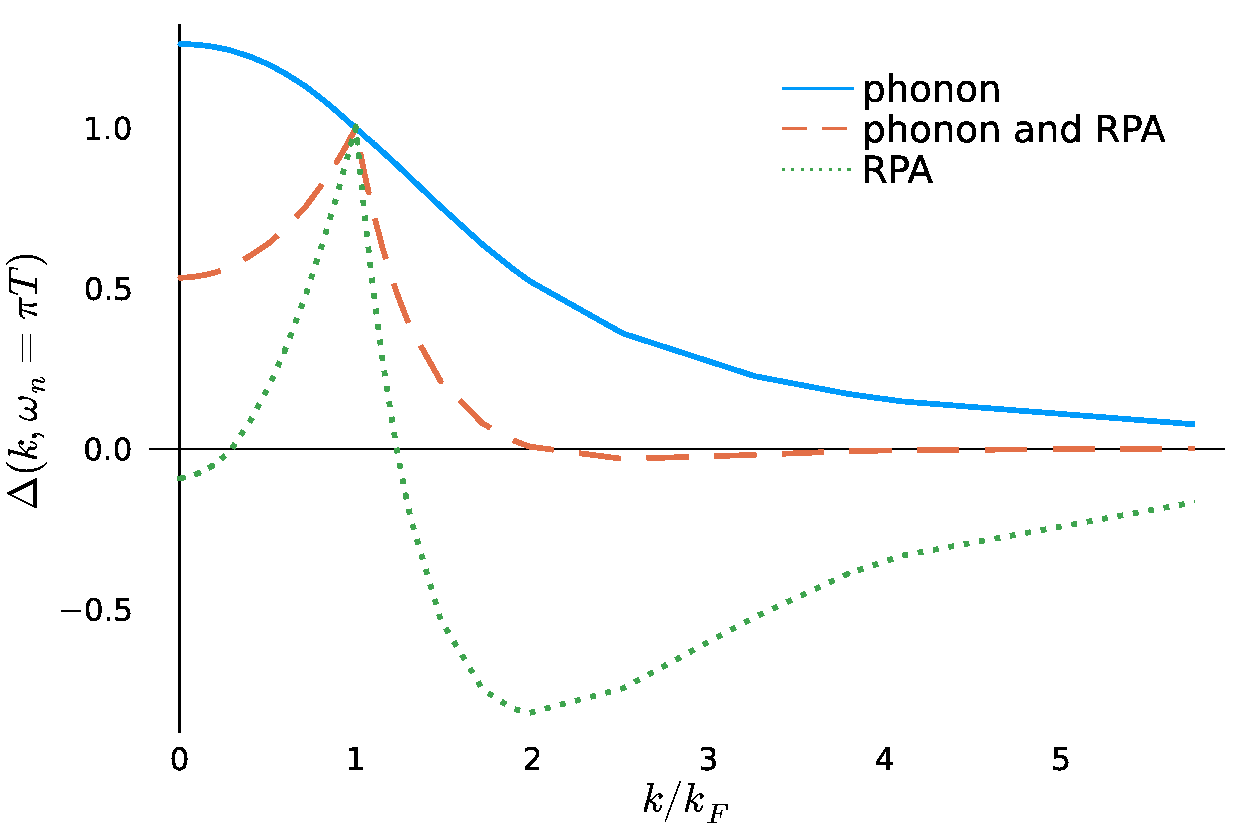}
\caption{\label{fig:delta} Gap function dependence on frequency at $k=k_F$(left panel) and  momentum at $\omega_{n} = \pi T$ (right panel) when $\Gamma$ is equal to either $\Gamma_{\rm ph}$ ,$\Gamma_{\rm RPA}$, or $\Gamma_{\rm ph}+\Gamma_{\rm RPA}$ at $r_s = 3$ [with $\Delta(k_F, \pi T)$ normalized to unity]. 
} 
\end{figure*} 

\textit{Results}.\textemdash
In Fig.~\ref{fig:main} we show the breakdown of the coupling constant into
$U$ and $f_{\rm FL}$ and the resulting values of $T_c$ for three different choices 
of $\Gamma$. We first consider the case when the Coulomb repulsion is omitted 
and $\Gamma $ is based on the electron-phonon interaction with $a = 0.8$   and $\omega_{\rm ph} = 0.01E_F$ in (\ref{eq:elph}). According to Migdal's theorem \cite{migdalINTERACTIONELECTRONSLATTICE}, the phonon-mediated vertex correction
is proportional to $a\omega_{\rm ph}/E_F$ and can be safely neglected. 
This calculation serves as a ``baseline" for examining effects induced by the 
Coulomb repulsion. For $\Gamma=\Gamma_{\rm ph} +\Gamma_{\rm RPA} $, the pseudopotential 
$U$ is first reduced to a minimum value at $r_s\approx0.75$ but then starts to increase and eventually surpasses the electron-phonon interaction value at $r_s \approx 2$ (cf. Ref.~\cite{rietschelRoleElectronCoulomb1983}). 
However, the Fermi liquid factor $f_{\rm FL}$ is getting progressively smaller with
increasing $r_s$. The net effect on the critical temperature is also non-monotonic, but the behavior of $T_c(r_s)$ is not as dramatic because the increase in $U$ at $r_s >1 $ is partially 
compensated by the suppression of $f_{\rm FL}$. 

When vertex corrections are accounted for Coulomb interaction and 
$\Gamma=\Gamma_{\rm ph} +\Gamma_{\rm KO}$, the Fermi liquid factor $f_{\rm FL}$ remains essentially the same for all values of $r_s$. However, changes in $U$ are relatively small (less than 20\%) only for 
$r_s < 1$. The most significant difference is the shift in the point of onset of the Coulomb enhancement of $U$:
from $r_s \approx 2$ 
to $r_s \approx 3$. This result underlines the importance of approximation-free  
high-order diagrammatic calculations. Nevertheless, the non-monotonic behavior of $T_c$ and $U$ is a robust effect based on the dynamic screening mechanism
that tends to make effective Coulomb interactions attractive at large $r_s$. 
It is completely overlooked in the Coulomb pseudopotential treatment.
 If  we take our value of $T_c$ at $r_s=2$ and try to reproduce it with the help of McMillan's formula, the phenomenological parameter $\mu^*$
ends up to being close to $0.08$. 

There exists yet another fundamental reason for non-additive effects when two pairing mechanisms are combined (even if 
$f_{FL}$ factors are accounted for exactly).
If $\lambda_{i=1,2}$ and $\Delta_i$ are the largest eigenvalue and its eigenvector for matrix $\Gamma_i$ and 
$\Delta_1 \ne \Delta_2$ then the largest eigenvalue of $\Gamma=\Gamma_1+\Gamma_2$ is \textit{always smaller} than
$\lambda_1+\lambda_2$ . In Fig.~\ref{fig:delta} 
we show gap function solutions for $\Gamma_1 = \Gamma_{\rm ph}$, 
$\Gamma_2=\Gamma_{\rm RPA}$ and $\Gamma=\Gamma_1+\Gamma_2$ at $r_s = 3$. 
One can see that the eigenvector ``mismatch" between these solutions is significant: 
While $\Delta_{\rm ph}$ is sign-positive and monotonic, $\Delta_{\rm RPA}$ changes sign both in the momentum domain and in the frequency domain and features a pronounced singularity at $k=k_F$.

\textit{Discussion and conclusion.}\textemdash 
It is instructive to put our findings in the context of historic developments. 
That Coulomb interaction can induce Cooper instability through dynamic screening mechanism has been known for decades. Early work \cite{Tolmachev1961} demonstrated that even if the Cooper channel coupling is repulsive at all 
frequencies, after its high-frequency part is renormalized to a 
smaller value the effective low-frequency potential might end up being attractive.
Later, Takada and others calculated critical temperatures of the UEG numerically 
using various approximate forms of the screened potential~ \cite{takadaSuperconductivityOriginatingRepulsive1989,takadaPwavePairingsDilute1993,rietschelRoleElectronCoulomb1983} featuring singular frequency/momentum  dependence (ignored without justification by introducing parameter $\mu$). 
These results raise an obvious question: Why are phenomenological values of $\mu^*$ used in material science always repulsive if Coulomb interaction alone can be the pairing glue? 

Several studies attempted to account for Coulomb effects on superconductivity beyond the Coulomb pseudopotential~\cite{richardsonEffectiveElectronelectronInteractions1997b,ludersInitioTheorySuperconductivity2005,weiHightemperatureSuperconductivityMonolayer2021}. Most relevant to our study is the work by Richardson and Ashcroft \cite{richardsonEffectiveElectronelectronInteractions1997b} who calculated $T_c$ 
for several metals by treating the electron-phonon, Eq.~(\ref{eq:elph}),
and Coulomb interactions on equal footing. They reported that in Lithium 
(with $r_s = 3.25$) the inclusion of Coulomb interaction leads to smaller $T_c$. 
Our results explain, that for large values of $r_s$ the suppression of 
$f_{\rm FL}$ is significant and cannot be dismissed as prescribed by the McMillan's formula. However, this fact was not well established at the time, and Richardson and Ashcroft tried to accommodate all effects into the framework of the existing phenomenological treatment.

By separating the Coulomb suppression of the Fermi liquid factor $f_{\rm FL}$ from its 
contribution to the low-frequency pseudopotential $U$, we shed light 
on the origin of the small critical temperatures observed experimentally when compared
with predictions of the Migdal-Eliashberg theory. 
We reveal that the Coulomb contribution to $U$ changes from repulsive to attractive
and conclude that the original interpretation of $\mu^*$ is incorrect and misleading in two ways: 
(i) The scenario of enhancement of attractive $U$ due to the dynamic nature of screening is ignored, leading to the false impression that $\mu^*$ is always positive; 
(ii) strong renormalization of Fermi liquid properties is ignored, while it can easily reduce the effective coupling by a factor of 2. These two mistakes partially compensate each other in the phenomenological treatment, yielding reasonable effective coupling constants $g$ within the freedom of choosing $\mu^*$. However, the actual 
microscopic picture behind the procedure is missed. 

The failure to appreciate the non-additivity of the phonon and Coulomb contributions to the effective coupling constant $g$---implied by the structure of Eq.~(\ref{g_U_f}) and also by the eigenvector mismatch (Fig.~\ref{fig:delta})---can lead to qualitatively wrong conclusions. 
For example, Ref.~\cite{rietschelRoleElectronCoulomb1983} stated that the
RPA is a deficient approximation at $r_s > 2$ because it predicts 
an attractive pseudopotential in contradiction with the ``experimentally established" $\mu^* > 0$. Taking proper account of all the aspects of the interplay between dynamically screened Coulomb repulsion and (alternative) pairing mechanisms may bring insights in the search for new superconducting materials, especially in cases when McMillan's equation fails qualitatively.

\begin{acknowledgements}

N.V.P., B.V.S., and T.W. acknowledge support by the National Science Foundation under Grant No. DMR-2032077. X.C. and K.C. acknowledge support from the Simons Collaboration on the Many Electron Problem. This work relies on the NUMERICALEFT package \cite{numericalEFT} , which is numerical toolbox dedicated for numerical effective field theory applications.

\end{acknowledgements}

%

\end{document}



\title{Supplemental Material\\ Origin of Coulomb Pseudopotential}
\author{Tao Wang}
\affiliation{Department of Physics, University of Massachusetts, Amherst, MA 01003, USA}
\author{Xiansheng Cai}
\affiliation{Department of Physics, University of Massachusetts, Amherst, MA 01003, USA}
\author{Kun Chen}
\email{kunchen@flatironinstitute.org}
\affiliation{Center for Computational Quantum Physics, Flatiron Institute, 162 5th Avenue, New York, New York 10010}
\author{Boris V. Svistunov}
\email{svistunov@physics.umass.edu}
\affiliation{Department of Physics, University of Massachusetts, Amherst, MA 01003, USA}
\affiliation{Wilczek Quantum Center, School of Physics and Astronomy, Shanghai Jiao Tong University, Shanghai 200240, China}
\author{Nikolay V. Prokof'ev}
\email{prokofev@physics.umass.edu}
\affiliation{Department of Physics, University of Massachusetts, Amherst, MA 01003, USA}

	
\date{\today}
\maketitle
\section{Symmetrized Discrete Lehmann representation}

Accurately reproducing singular properties of the imaginary-time Green's function, $G(\tau)$, at low temperature $T=\beta^{-1}$ is a challenging numerical problem.
The recently developed Discrete Lehmann representation (DLR) \cite{kayeDiscreteLehmannRepresentation2022} offers a solution in the form of a compact ansatz for $G(\tau)$. Here we briefly render DLR in its original form \cite{kayeDiscreteLehmannRepresentation2022} along with a generalized version---symmetrized DLR (SDLR) proposed by some of us \cite{SDLR}. The SDLR approach is particularly convenient when working in the Matsubara-frequency representation (and demonstrates other notable advantages that are discussed elsewhere \cite{SDLR}).

The standard spectral (Lehmann) repbresentation for $G$ reads (here and below we do not show the dependence on momentum and spin variables):
\begin{equation}
G(\tau)=\int_{-\infty}^{\infty} K(\tau, \omega) \rho(\omega) d \omega \, ,
\label{Lehmann}
\end{equation}
where $\omega$ is the real frequency and $\rho(\omega)$ is the spectral density---a real-valued non-negative function. The convolution kernel $K(\tau, \omega)$ is a universal function that only depends on the particle statistics and temperature. This universality allows one to interpret (\ref{Lehmann}) as 
a linear superposition of fixed functions $K(\tau, \omega)$, with $\omega$ labelling the functions and $\rho(\omega)$ playing the role of the 
superposition  coefficients.

The key observation behind DLR and similar approaches is that the ``vectors"
$K(\tau, \omega)$ are massively correlated. As a result, the {\it effective dimension} of the functional space spanned by the set $K(\tau, \omega)$ with realistic $\rho(\omega)$'s  turns out to be as small as
\begin{equation}
    r \sim \log \frac{E_{uv}}{T} \log \frac{1}{\epsilon}\, ,
    \label{Eq:degree}
\end{equation}
where $E_{uv}$ is the frequency cutoff beyond which the physical spectral density can be set to zero and $\epsilon$ is the specified accuracy with which $G$ is reproduced. The efficient representation of the Green's function can then be achieved by identifying a discrete set of $r$ optimal basis functions. According to (\ref{Eq:degree}), the number of basis functions 
increases very slowly with decreasing $T$ and $\epsilon$, and in practice only several dozens of basis functions are needed.

\subsection{Discrete Lehmann Representation}

Within the DLR approach  \cite{kayeDiscreteLehmannRepresentation2022}, the basis functions $K(\tau,\omega_k)$ correspond to $r$ optimally chosen frequency points $\omega_k$  ($k=1, 2, ..., r$) identified by a pivoted QR algorithm \cite{kayeDiscreteLehmannRepresentation2022}. The Green's function is then represented as
\begin{equation}
G(\tau) \approx G_{\mathrm{DLR}}(\tau) \equiv \sum_{k=1}^{r} K\left(\tau, \omega_{k}\right) \widehat{\rho}_{k} \, ,
\label{Eq.DLR}
\end{equation}
i.e. the spectral density is replaced with $r$ ``quasiparticle poles.''
For a given $G(\tau)$ data set, the pole residues $\widehat{\rho}_{k}$
are obtained from the least-squares fitting. (In practice, the 
fitting protocol is naturally implemented in the Matsubara-frequency representation; see below.)

The Green's function $\tau$-dependence on the $[-\beta, \beta]$ interval is unambiguously fixed by its behavior on the $(0, \beta)$ interval. Within this interval, the kernel $K(\tau,\omega)$ can be rendered the same for bosons and fermions [by absorbing statistics-dependent but $\tau$-independent factors into the ``auxiliary" spectral density $\rho(\omega)$]:
\begin{equation}
K(\tau, \omega) = e^{-\omega \tau}\, ,\qquad  \tau \in (0, \beta) \, .
\label{Eq:fermiDLR}
\end{equation}

While having simple exponential basis functions in the $\tau$-representation is convenient, the least-squares fitting procedure for obtaining $\widehat{\rho}_{k}$
is most naturally implemented in the Matsubara-frequency representation because in diagrammatic calculations $G$ is extracted from the Dyson equation solved at frequencies $\xi_m=2m \pi T $ for bosons and $\xi_m=(2m+1)\pi T$ for fermions 
($m=0,\pm 1,\pm 2, \ldots$). The two representations are related by
\begin{equation}
 G(\xi_m)= \int_0^{\beta} e^{i\xi_m \tau}\, G(\tau) \, d\tau \, .
 \label{Green_xi}
\end{equation}
After $G(\xi_m)$ is computed for a pre-selected finite set of frequencies 
the data is used for the least-squares fitting:
\begin{equation}
\sum_{k=1}^{r} K\left(\xi_m, \omega_{k}\right) \widehat{\rho}_{k} \, \approx \, G(\xi_m) \, .
\label{fitting_freq}
\end{equation}
Here $K\left(\xi_m, \omega \right)$ is the Fourier transformed kernel  (\ref{Eq:fermiDLR}):
\begin{equation}
K(\xi_m, \omega) = -\frac{1\pm e^{-\omega \beta}}{i\xi_m-\omega} \, ,
\end{equation}
with $\pm$ sign for fermions/bosons. 
 
\subsection{Symmetrized Discrete Lehmann Representation} 


We now discuss the Symmetrized Discrete Lehmann Representation based on the generic decomposition of $G(\tau)$ into the sum of particle-hole symmetric, $G^+(\tau)$, and particle-hole anti-symmetric, $G^-(\tau)$, parts:
\begin{equation}
G(\tau) = G^{+}(\tau) + G^{-}(\tau)\, ,
\label{decomp}
\end{equation}
where  
\begin{equation}
G^{\pm}(\tau) = \frac{G(\tau) \pm G(\beta-\tau)}{2} \, ,
\label{G_plus_minus}
\end{equation}
are obeying $G^{\pm}(\tau) = \pm G^{\pm}(\beta-\tau)$. 
Despite the fact that {\it exact} Lehmann representations for the two functions share the same spectral density $\rho(\omega)$, 
\begin{equation}
G^{\pm}(\tau)=\int_{-\infty}^{\infty} K^{\pm}(\tau, \omega) \rho(\omega) d \omega \, , \label{Eq:SymLehmann1}
\end{equation}
\begin{equation}
K^{\pm}(\tau,\omega) = \frac{K(\tau,\omega) \pm K(\beta-\tau,\omega)}{2} \, ,
\label{Eq:SymLehmann2}
\end{equation}
the {\it discrete} approximations are not supposed to be 
strictly related. Indeed, for discrete sums 
\begin{eqnarray}
G^{\pm}_{\mathrm{SDLR}}(\tau) \equiv \sum_{k=1}^{r}
K^{\pm}\left(\tau, \omega^{\pm}_{k}\right) \widehat{\rho}^{\pm}_{k} \, ,
\label{Eq:SDLR}
\end{eqnarray}
the two sets of optimised frequencies, $\omega^{\pm}_{k}$, obtained by the algorithm of Ref.~\cite{kayeDiscreteLehmannRepresentation2022},
are not identical. Correspondingly, the basis functions and pole residues, $\widehat{\rho}^{\pm}_k$, also turn out to be different.  

The utility of working with $G^+$ and $G^-$ comes form the following properties:
\begin{equation}
G^+(\xi_m)=\left\{
\begin{array}{l}
{\rm Re}\,  G(\xi_m)\, , ~\qquad \mbox{bosons} \; , \\
i\, {\rm Im}\,  G(\xi_m)\, , \qquad \mbox{fermions} \; ,\end{array} \right.  
\label{G_plus_xi}
\end{equation}
\begin{equation}
G^-(\xi_m)=\left\{
\begin{array}{l}
  i\, {\rm Im}\,  G(\xi_m)\, , \qquad \mbox{bosons} \; , \\
 {\rm Re}\, G(\xi_m)\, , ~~ \qquad\mbox{fermions} \; ,\end{array} \right. 
\label{G_minus_xi}
\end{equation}
\begin{eqnarray}
K^{+}(\xi_m, \omega) &=& \frac{2i\xi_m}{\omega^2+\xi_m^2} (1+e^{-\omega \beta}) \quad \mbox{(fermions)} \, , ~~\\
K^{-}(\xi_m, \omega) &=& \frac{2\omega}{\omega^2+\xi_m^2} (1+e^{-\omega \beta})\quad \mbox{(fermions)} \, ,~~ \\
K^{+}(\xi_m, \omega) &=& \frac{2\omega}{\omega^2+\xi_m^2} (1-e^{-\omega \beta}) \quad \mbox{(bosons)} \, ,\\
K^{-}(\xi_m, \omega) &=& \frac{2i\xi_m}{\omega^2+\xi_m^2} (1-e^{-\omega \beta})\quad \mbox{(bosons)} \, .
\end{eqnarray}
It is easy to see that the problem of extracting the weights $\widehat{\rho}^{\pm}_k$ from $G(\xi_m)$ splits into two independent real-valued problems, independently utilizing the real and imaginary parts of $G(\xi_m)$.

We found that the original DLR scheme suffers from instabilities when the Dyson and self-energy equations are iterated, e.g. when 
implementing the so-called $GW_0$ approximation. The SDLR scheme
is free of this problem. More details are provided in Ref.~\cite{SDLR}
 
We implemented SDLR in a julia package open-sourced in the repository \url{https://github.com/numericalEFT/Lehmann.jl}. We also calculated the basis functions for various temperatures and accuracy bounds, and published them on \url{https://github.com/numericalEFT/Lehmann.jl/tree/main/basis}.